\documentclass[12pt]{article} 
\usepackage{ol2}
\usepackage[draft]{hyperref}
\usepackage{amsmath}

\begin{document}


\title{Continuum generation by dark solitons}


\author{C. Mili\'an$^{1,2}$, D.V. Skryabin$^{1,*}$, A. Ferrando$^3$}

\address{
$^1$Centre for Photonics and Photonic Materials, Department of
Physics, University of Bath, Bath BA2 7AY, UK
\\ $^2$Instituto ITACA, Universidad Polit\'ecnica
de Valencia, Camino de Vera S/N, 46022 Valencia, Spain\\
$^3$Departamento de \'Optica, Interdisciplinary Modeling Group InterTech,
Universidad de Valencia, Dr. Moliner 50, 46100 Burjassot, Valencia, Spain
 \\$^*$Corresponding author:
d.v.skryabin@bath.ac.uk}

\begin{abstract}
We demonstrate that  the dark soliton trains in optical fibers with
a zero of the group velocity dispersion can generate broad
spectral distribution (continuum) associated with the resonant
dispersive radiation emitted by  solitons. This radiation is
either enhanced or suppressed by  the Raman scattering depending on
the sign  of the third order dispersion.
\end{abstract}



\noindent
One of many fundamental and practical outcomes of
research into supercontinuum generation in optical fibers
\cite{Dudley2006,Knight2007a} has been a change of emphasis in studies of temporal
solitons. If in the pre-supercontinuum era fiber solitons were mostly
perceived as information carriers \cite{Mollenauer2006}, now days they are widely recognized
and researched as  facilitators and mediators of efficient
frequency conversion in infrared and visible parts of spectrum
\cite{Dudley2006,Knight2007a}.

Trains of bright solitons  naturally emerge in fibers
from an intense pump pulse  and their role in supercontinuum generation
and frequency conversion has been extensively explored \cite{Dudley2006,Knight2007a}.
In particular, it has been demonstrated that the interplay of  the Raman scattering
and third order dispersion (TOD) can lead to the exponential amplification of the
resonant (Cherenkov) radiation emitted by a soliton \cite{Skryabin2003}. This happens if the central
soliton frequency is shifted by  the Raman effect towards a zero of the
group velocity dispersion (GVD), which is possible if TOD is negative \cite{Skryabin2003}.
Amplification of the resonant radiation
has been used to create  coherent infrared   fiber based sources \cite{Falk2008,Liu2008,Zhigang2009}.
While, the widest octave broad supercontinua extended from $\sim$400nm to $\sim$2$\mu$m have
been most commonly observed across the zero GVD point with  positive TOD \cite{Ranka2000}.
Generation of these spectra is fueled by the cascaded scattering of the short
wavelength radiation on and its trapping by the Raman
solitons  \cite{Knight2007a,Gorbach2007a,Cumberland2008}.

Theory of the radiation emission by bright solitons has been developed
well before the recent wave of supercontinuum research  \cite{Dudley2006,WAI1990}.
However, importance of  the Raman effect  in the soliton-radiation interaction has
been revealed and explained only recently \cite{Skryabin2003,Gorbach2007a}. Similarly, it is
known  that {\em dark solitons} also emit radiation when perturbed by TOD
\cite{KARPMAN1993,Afanasjev1996}, but the interplay of this process
with the Raman scattering has not been studied so far and is the subject of this Letter.
Trains of dark solitons can  be created with the established methods
\cite{ROTHENBERG1991,ROTHENBERG1992} and, as we will demonstrate below, can generate continuum
of dispersive radiation.
Continuum generation by dark solitons can be important not only from a fundamental point
of view, but also practically relevant if the pump frequency is
in the normal GVD range of a fiber and  generation of a
broadband signal across the zero GVD wavelength is sought.

A model we use is the dimensionless NLS equation with the TOD and Raman terms
\begin{eqnarray}
\nonumber &&i\partial_{z} A={1\over 2}\partial_{t}^2 A+i\epsilon\partial_{t}^3
A \\&& - (1-\theta)|A|^2 A-\theta A\int R(t')|A(t-t')|^2dt'.\label{d1}
\end{eqnarray}
Here $R$ is the Raman response function of silica and $\theta=0.18$ \cite{Dudley2006}.
$t$ is the delayed time measured in the units of   $T=100$fs.
$\epsilon=\beta_3/[6T\beta_2]$ is a small parameter characterizing the ratio between TOD
($\beta_3$) and normal GVD ($\beta_2>0$). $A$ is
the amplitude of the electric field scaled to $\sqrt{P}$, where
$P=1/[\gamma L_D]$. $L_D=T^2/\beta_2$ is the dispersion length and
$\gamma$ is the nonlinear fiber parameter. Assuming $A\sim e^{-i\delta t}$,
the zero GVD frequency is $\delta_0=-1/[6\epsilon]$. For
 positive (or negative) TOD the anomalous GVD range hostile to the existence of dark solitons
is located for  $\delta<\delta_0$ (or $\delta>\delta_0$).
A unit of $\delta$ corresponds to $10$THz.

We are interested in $A$ consisting of the dark soliton $F$
plus small radiation $g$: $A(z,t)=(F(\tau,z)+g(\tau,z))e^{i\kappa^2z}$,
where $\tau=t-z\kappa\sin\phi $.
The soliton solution is  \cite{Kivshar1998}
\begin{equation}
F=\kappa ~{\rm tanh}[\tau\kappa\cos\phi]\cos\phi-i\kappa\sin\phi. \label{d3}
\end{equation}
$\kappa>0$ is the amplitude of the soliton background.
$\phi$ is  the darkness parameter, with the condition $\cos\phi\ge 0$ implying  $\phi\in [-\pi/2,\pi/2]$.
$\phi=0$ corresponds to the black soliton having zero amplitude in the middle
and propagating with the group velocity matching the background velocity \cite{Kivshar1998}.
$\phi\in (0,\pi/2)$ and  $\phi\in (-\pi/2,0)$ correspond
to the dark solitons propagating, respectively, slower and faster than the background wave.

Solutions of a linearized equation for $g$ are sought
in the form \begin{equation}
g=G_1e^{i\delta_r\tau-i\lambda z-i\phi}+G_2^*e^{-i\delta_r\tau+i\lambda
z+i\phi}.\label{d6}
\end{equation}
Note, that for $A\sim e^{-i\delta t}$, the exponent
$e^{i\delta_r\tau}$ with $\delta_r>0$ corresponds to the overall negative frequency shift.
Condition $\lambda=0$ leads to the resonant frequency \cite{KARPMAN1993,Afanasjev1996}
\begin{equation}
\label{delta}  \delta_r^2= {1\over 2\epsilon^2}
[\alpha +\sqrt{\alpha^2+4\kappa^2\epsilon^2\cos^{2}\phi}],~\alpha\equiv 2\epsilon\kappa\sin\phi
+{1\over 4}.\end{equation}
Expanding in $\epsilon$ we find $\delta_r=1/(2\epsilon)(1+4\kappa\epsilon\sin\phi+2\kappa^2\epsilon^2\cos^2\phi+O(\epsilon^3))$,
so that $|\delta_r|>|\delta_0|$, i.e. the resonant frequency is detuned further away from the soliton,
than the zero GVD frequency.
An important difference with  bright solitons is that  the amplitudes $G_1$
and $G_2$ are not independent here, but are coupled by
nonlinear interaction mediated by the finite amplitude soliton background.
The ratio of these amplitudes works out as
\begin{equation}
{G_1\over G_2}=-{1\over \epsilon^2\kappa^2}\left[{1+4\kappa\epsilon\sin\phi+O(\epsilon^2)\over 1+7\sin^2\phi}\right].
\label{amp}\end{equation}

Eq. (\ref{d6}) implies that the radiation spectrum is expected to peak on both sides
from the soliton spectral center at $\delta=0$. Since $|\delta_r|>|\delta_0|$, one of
the radiation peaks  belongs to the normal GVD range and the other one to the anomalous
GVD.   From Eq. (\ref{amp}) it follows
that the amplitude of the radiation belonging to the anomalous GVD range is  much
larger. The above holds for both signs of TOD, i.e. $\epsilon$.
Thus the radiation is primarily emitted into a spectral
range where the balance of GVD and nonlinearity is not able
to support dark solitons.
Figs. 1(a,b) show solitons and their radiation using  XFROG spectrograms \cite{Dudley2006}
calculated with the Raman effect
disregarded ($\theta=0$). For $\epsilon>0$ (Fig. 1(a)) and $\epsilon<0$ (Fig. 1(b))
the strong and weak radiation peaks are swapped, so that  the strong  one
always belongs to the anomalous GVD range. Spectrograms used here are computed as
$S(\delta,t)=|\int A(t'){\rm sech}(t'-t)\exp[-i\delta t']dt'|^{2}$.
For  $\kappa\cos\phi=1$ and $t=0$    the dark soliton spectrogram, $A=F$, can be calculated explicitly
\begin{equation}
S\propto (\delta^2\cos^2\phi+2\delta\sin\phi\cos\phi+\sin^2\phi)~{\rm sech}^{2}[\pi\delta/ 2].
\label{spec}\end{equation}

Taking account of the Raman scattering significantly changes the radiation pattern, cf. Fig. \ref{fig1}
and Fig. \ref{fig2}. The Raman effect enhances both radiation peaks
for positive TOD and suppresses them for negative TOD. Physics behind this observation is as follows.
Spectrogram  $S(\delta,0)$ of the ideal black, $\phi=0$, soliton  consists of the two symmetric
in $\delta$  peaks (not clearly seen in Fig. 1, but obvious from Eq. (\ref{spec})).
It means that the resonant radiation
in the anomalous GVD range is excited by the black soliton
with equal efficiency for either positive or negative TOD.
The Raman scattering however transfers more energy into the lower frequency ($\delta<0$)
side of the soliton spectrum. It makes
more of the soliton  energy to flow into the anomalous GVD range, providing TOD is positive.
For negative TOD, however, the effective spectral center of mass
of the soliton is shifted by  the Raman effect
further into the normal GVD range, where little disturbs the balance of normal GVD and nonlinearity.
Naturally, this  depletes the energy transfer into radiation. Asymmetry
of the dark soliton  spectra  induced by the Raman effect assumed in the above discussion
can be readily verified numerically and was experimentally measured
20 years  ago \cite{WEINER1989}.

The soliton spectral asymmetry  induced by  the Raman scattering  can be mimicked
by taking the fast dark solitons ($\phi<0$), see Eq. (\ref{spec}).
Equally it is true to say, that the Raman scattering accelerates dark
solitons  by inducing the adiabatic drift of $\phi$ in the direction of  $-\pi/2$. Thus by taking
initially a faster (slower) soliton in a positive (negative) TOD fiber one shifts more of the soliton
energy to the anomalous GVD range, thereby  boosting  energy transfer into
dispersive radiation. This point is further discussed and illustrated below.

A train of dark solitons is an obvious candidate
for generation of multiple radiation peaks forming a continuous spectrum.
One of the straightforward ways to generate these trains is through the dispersion induced
interference of two suitably delayed short pulses \cite{ROTHENBERG1991,ROTHENBERG1992}.
Each interference dip  has its own carrier frequency associated with
the chirp developing across the whole interference pattern. The characteristic
GVD and nonlinear lengths can be made approximately equal and are
typically shorter then the TOD length. Thus one can  arrange initial conditions
so that the solitons are formed first and then the TOD induced radiation
starts to develop.

Fig. \ref{fig3}  shows time domain evolution of the soliton train
for $\epsilon>0$ (a) and for $\epsilon<0$ (b). Both are with the Raman effect included.
The solitons
are separated into two groups. The ones found for $t<0$ and $t>0$ are the fast ($\phi<0$)
and slow ($\phi>0$) solitons, respectively. For $\epsilon>0$ the spectra of
the fast solitons are shifted towards the range of anomalous GVD. Therefore these solitons are
more keen to emit radiation, than the slow solitons
(see pronounced oscillations developing for $t<0$ in Fig. \ref{fig3}(a)).
The Raman effect  enhances the soliton spectral asymmetry further and
boosts the radiation emission by the fast solitons, see Fig. \ref{fig3}(b). For $\epsilon<0$
spectra of the slow solitons have greater overlap with the anomalous GVD range. However, the Raman
effects tends to compensate for this. Thus the net effect is the radiation depletion,
cf. Fig. \ref{fig3}(a) and (b).

Spectral evolution along the fiber length
for the cases of positive and negative TOD corresponding to the time plots
in Fig. 3 are shown in Fig. \ref{fig5}. Far stronger  energy transfer
from the normal to the anomalous GVD range is
obvious in the positive TOD case.
Quantitative characterization of this effect is facilitated by Fig. \ref{fig4},
where the full lines
show the spectral distributions (plotted using the log-scale)
corresponding to the final time domain signals in Fig. \ref{fig3}.
The spectrum generated in the anomalous GVD range is few orders of magnitude stronger,
than the one in the negative TOD case.
Contrary, if the Raman effect is switched off, then
$\epsilon>0$ case is a mirror image of the $\epsilon<0$ case, cf. the dashed lines in
Fig. \ref{fig4}(a) and (b).

In summary, we have demonstrated that  the Raman scattering  significantly amplifies
resonant dispersive radiation emitted  by a train of dark solitons in fibers with the positive
third order dispersion. This leads to generation of a broad band continuum in the anomalous GVD
range, when a pair of pump pulses is applied in the normal GVD. Practical realizations of this
effect and its applications for frequency conversion  should be  investigated further.

CM gratefully thanks the Formación de Profesorado Universitario predoctoral grant.
DVS is grateful to A. Yulin and J. Dudley for early inputs.

\newpage

Figure captions

\begin{figure}[htb]
\caption{(Color online) XFROG spectrograms showing resonant radiation by a dark soliton without
the Raman effect ($\theta=0$).
Dashed vertical lines indicate $\pm\delta_r$ predicted by Eq. (\ref{delta}).
Full vertical lines indicate the zero GVD frequency $\delta_0$.
(a) $\epsilon=0.0833$ and (b) $\epsilon=-0.0833$. Other parameters are $\kappa=1$, $\phi=0$, $z=30$.
}
\label{fig1}\end{figure}

\begin{figure}[htb]
\caption{(Color online) The same as  Fig. \ref{fig1}, but with the Raman effect ($\theta=0.18$).
Strong emission of radiation by the black soliton, see  (a), is accompanied by creation of a shallow dark soliton with
$\phi$ close $-\pi/2$ (marked with an arrow).}
\label{fig2} \end{figure}

\begin{figure}[htb]
 \caption{(a) Time domain evolution of the dark soliton train in a fiber with positive TOD:
$\epsilon=0.0217$, $\theta=0.18$. Initial condition is  $A=\sqrt{10}[sech(t-3)+sech(t+3)]$.
 (b) is the same as (a), but for negative TOD: $\epsilon=-0.0217$.}
\label{fig3} \end{figure}

\begin{figure}
\caption{
(Color Online)
Spectral development of continuum generation by a train of dark solitons
corresponding to the time domain evolution in Fig. 3,
but for larger $z$. Full vertical lines indicate the zero GVD frequency.}
\label{fig5}\end{figure}

\begin{figure}
\caption{Full lines in (a) and (b) show spectra corresponding to the $z=10$ time domain signals
 in Fig. \ref{fig3}. Dashed lines
 show the corresponding spectra computed  without the Raman effect ($\theta=0$).}
\label{fig4}\end{figure}

\newpage

\setcounter{figure}{0}
\begin{figure}[htb]
\centerline{
\includegraphics[width=3.5cm]{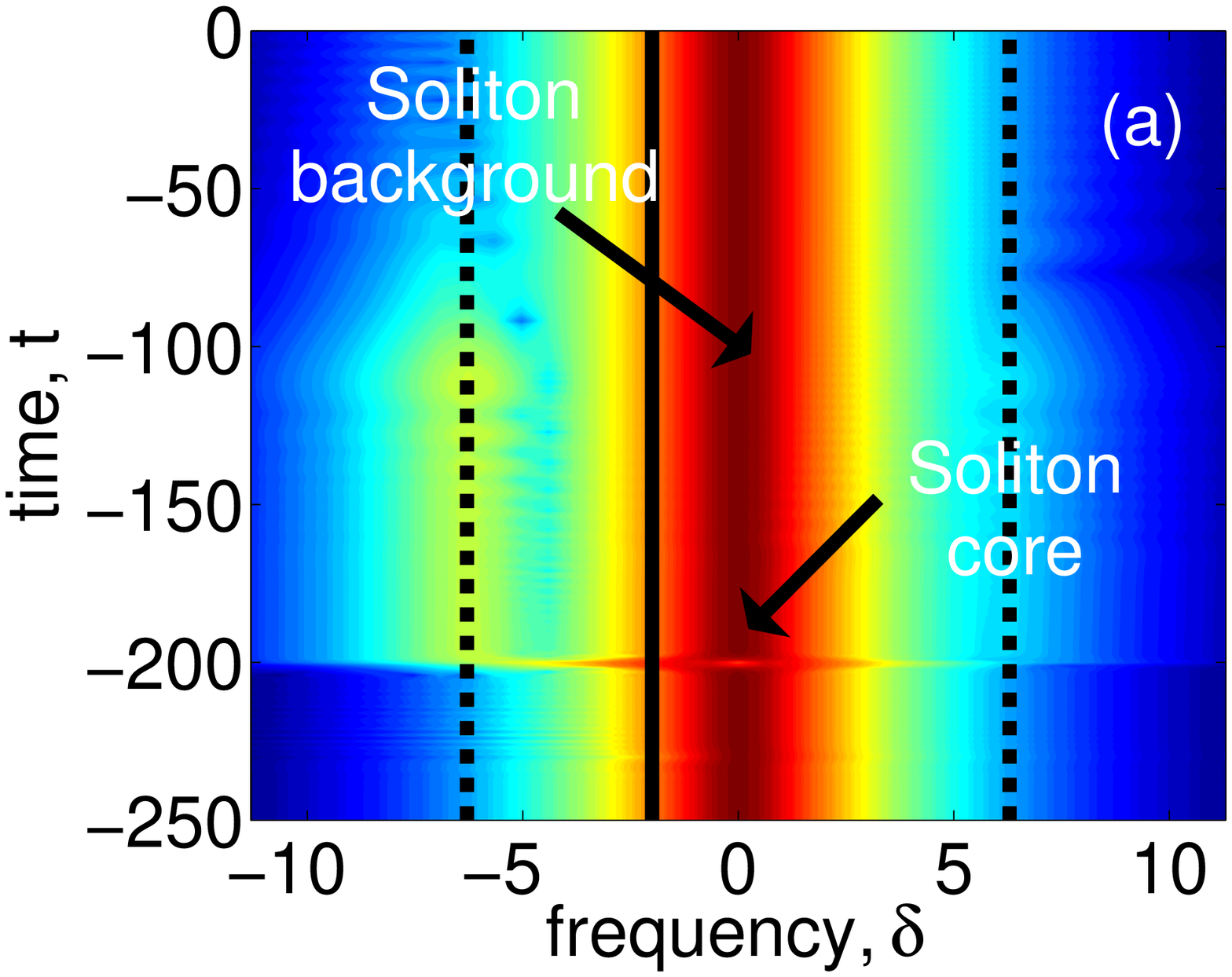}
\includegraphics[width=3.5cm]{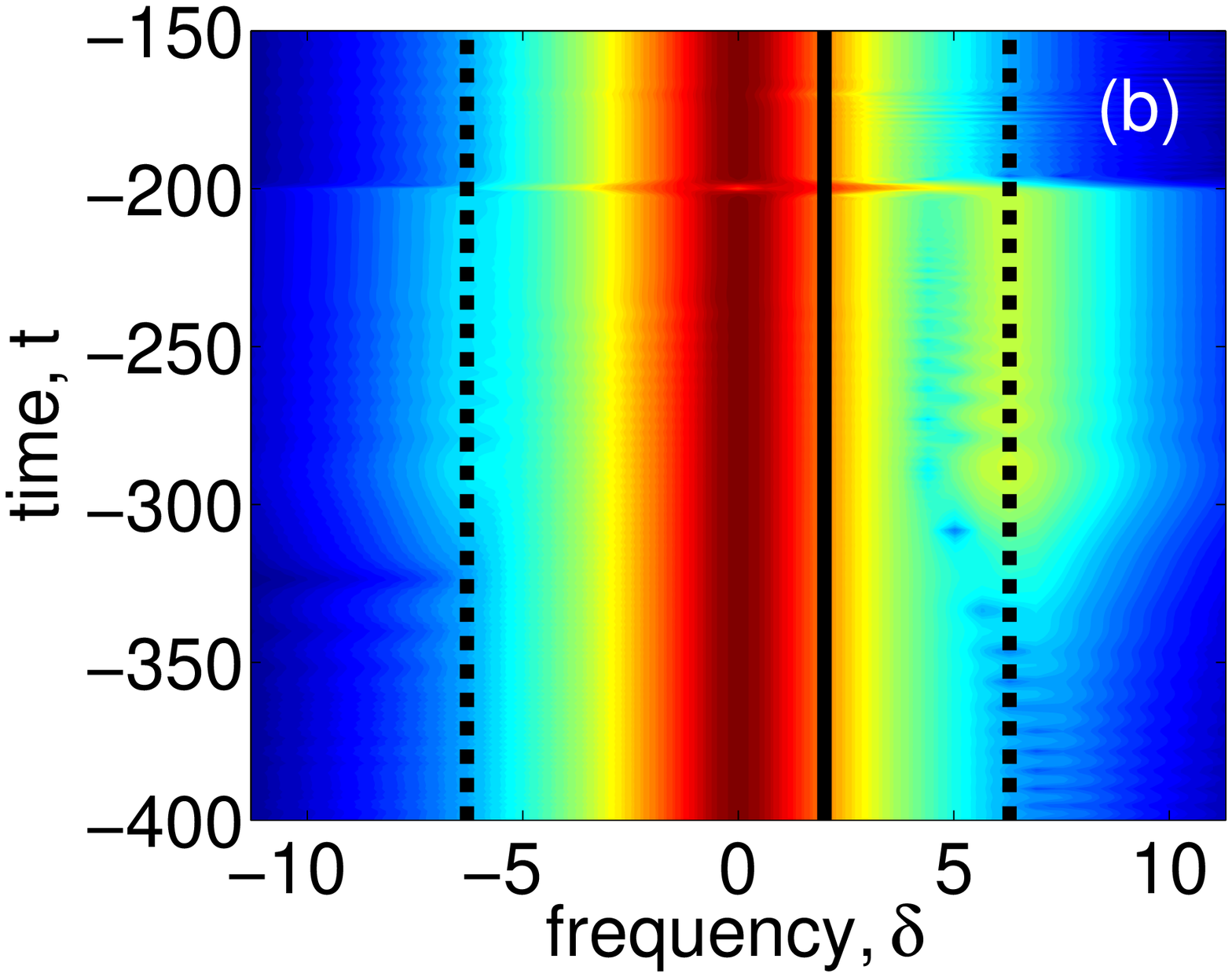}}
\caption{ }
\label{fig1}\end{figure}

\begin{figure}[htb]
\centerline{
\includegraphics[width=3.5cm]{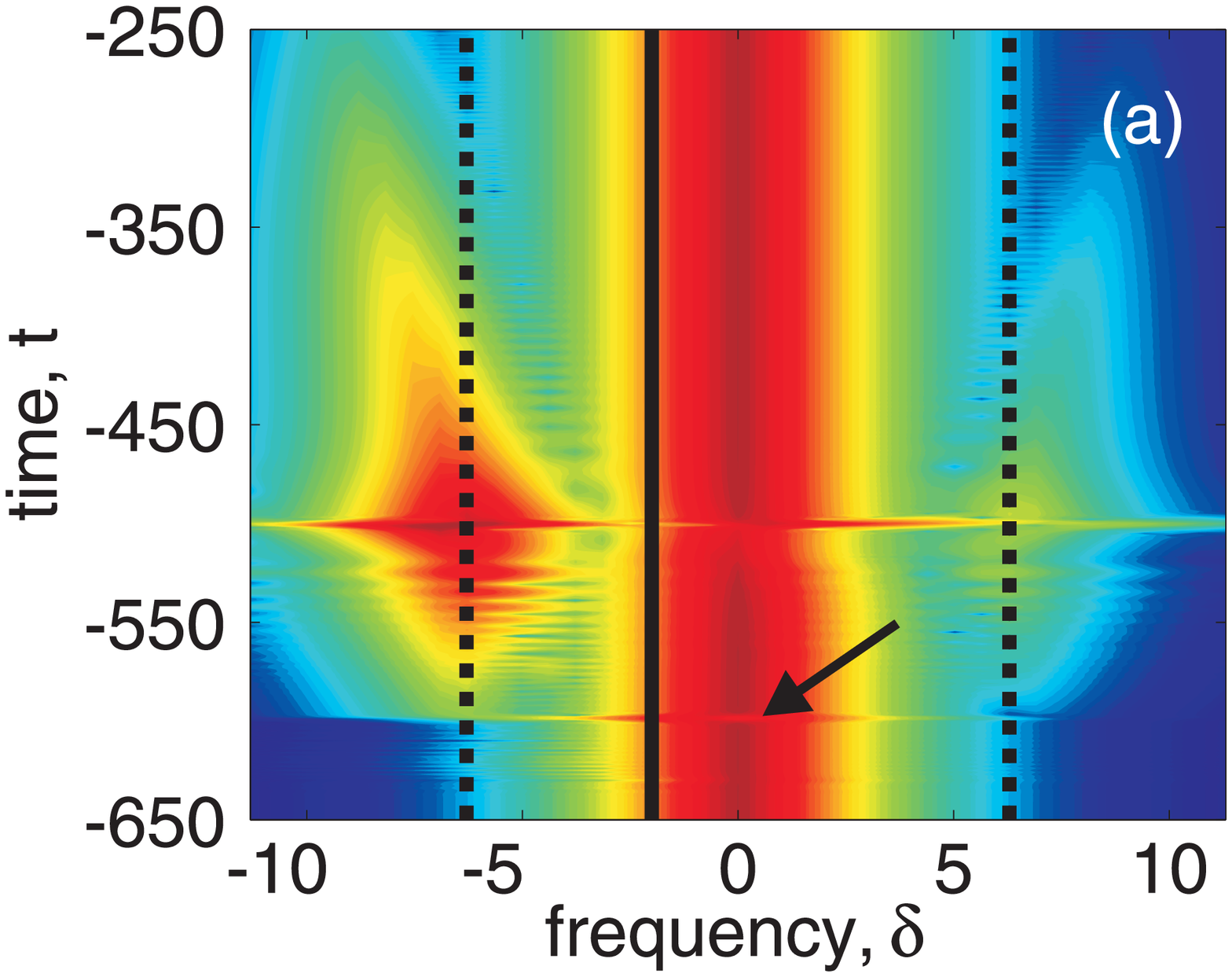}
\includegraphics[width=3.5cm]{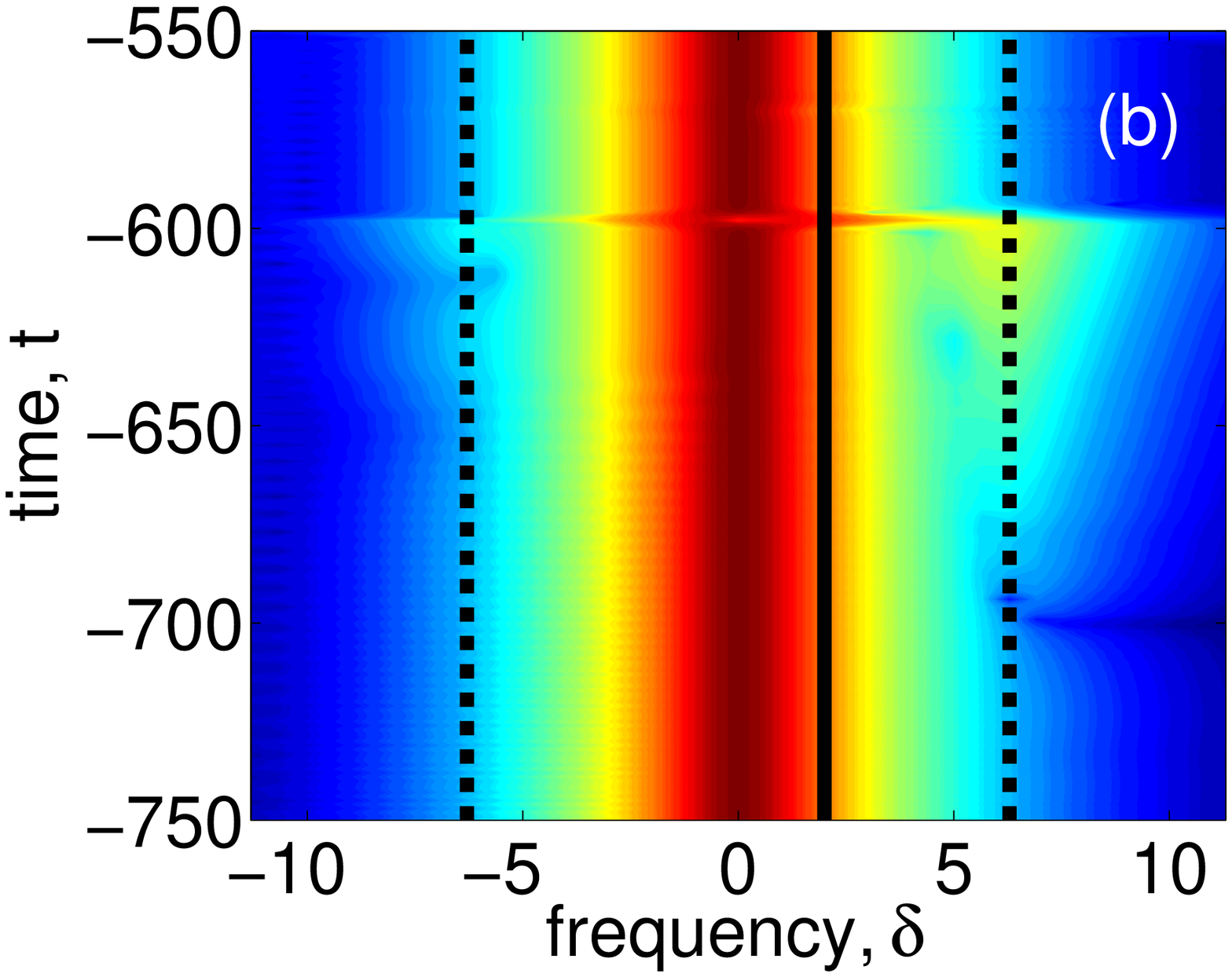}}
\caption{ }
\label{fig2} \end{figure}

\begin{figure}[htb]
\centerline{
\includegraphics[width=4.cm]{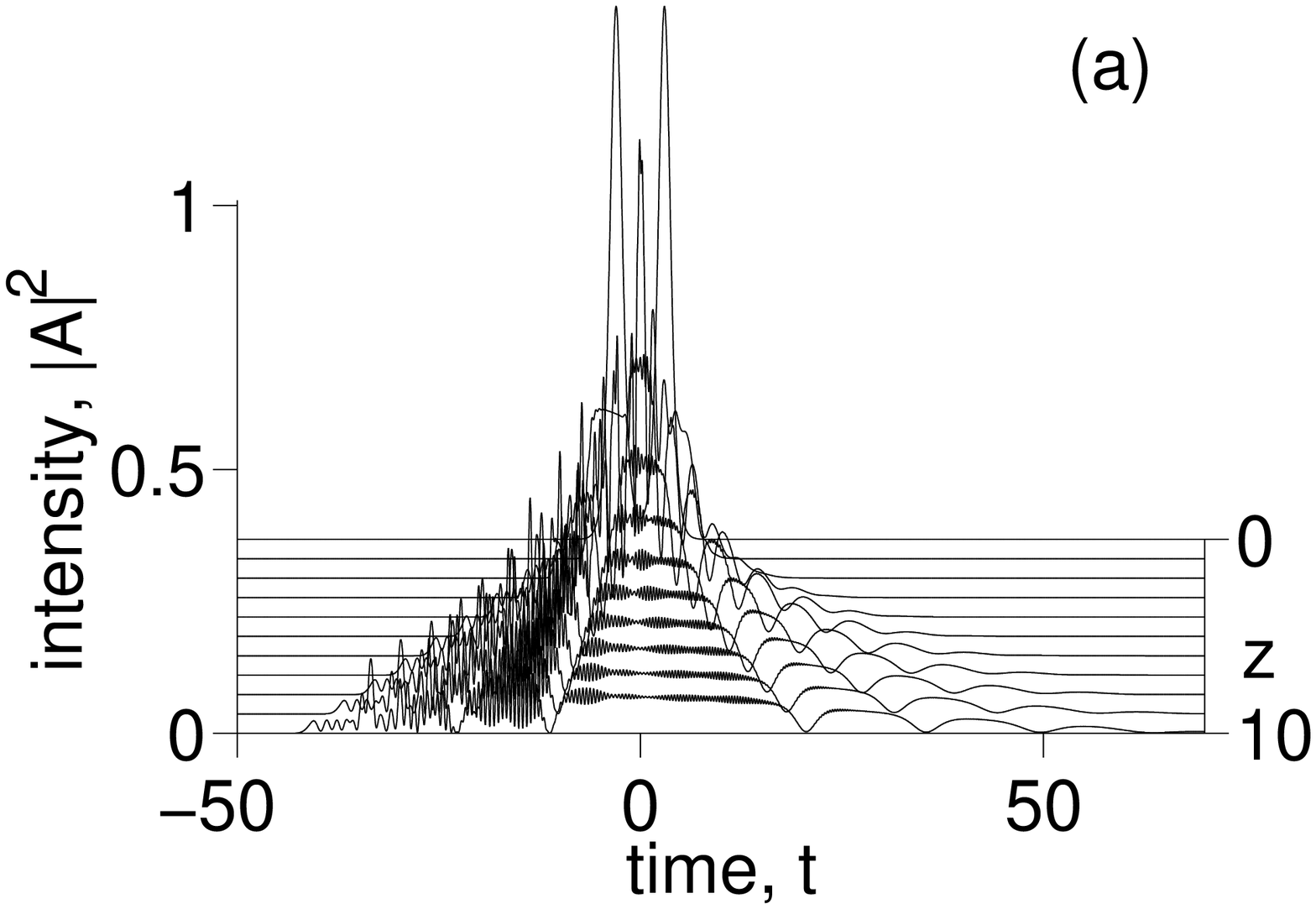}
\includegraphics[width=4.cm]{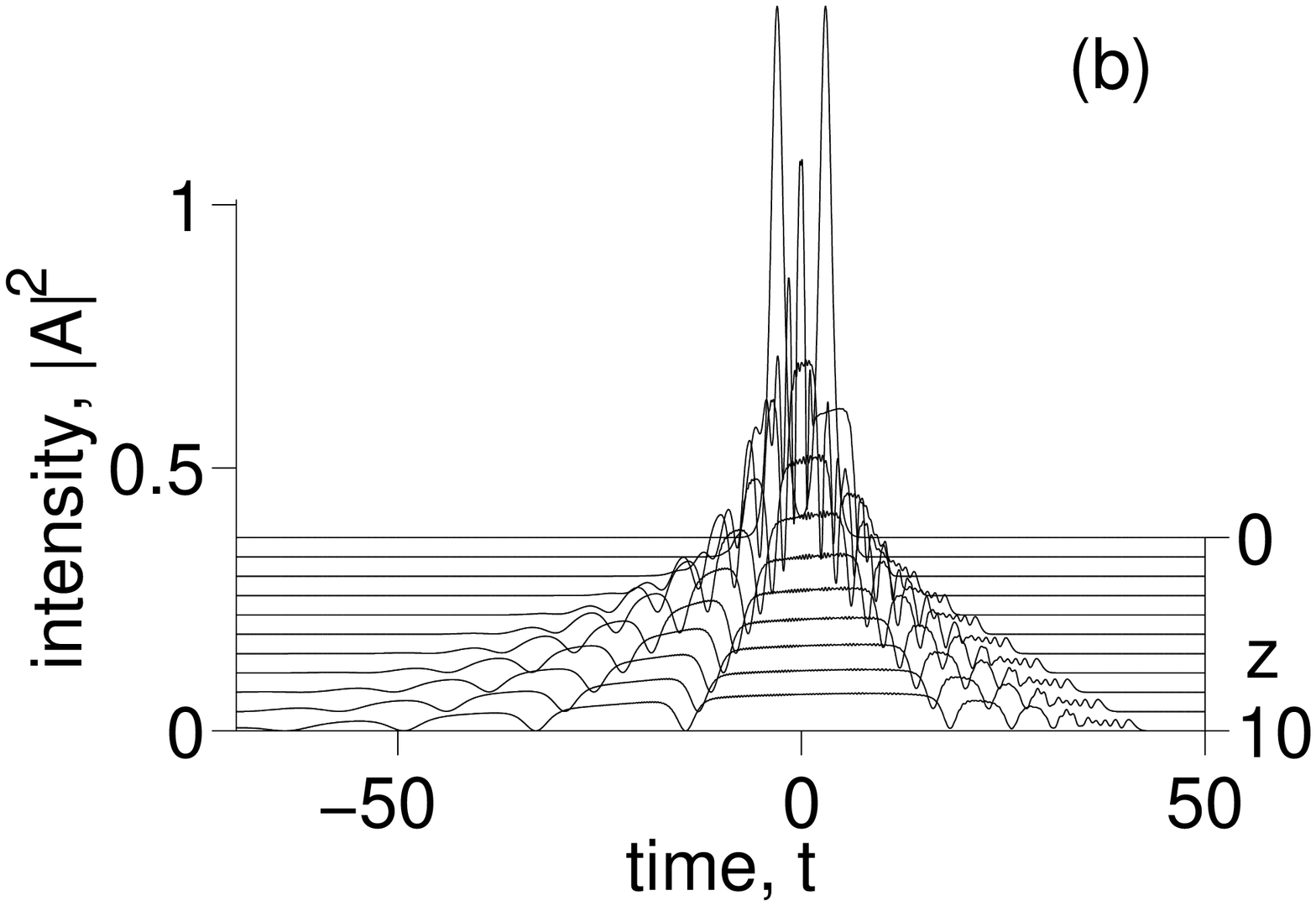}
}
 \caption{ }
\label{fig3} \end{figure}

\begin{figure}
\centerline{
\includegraphics[width=3.5cm]{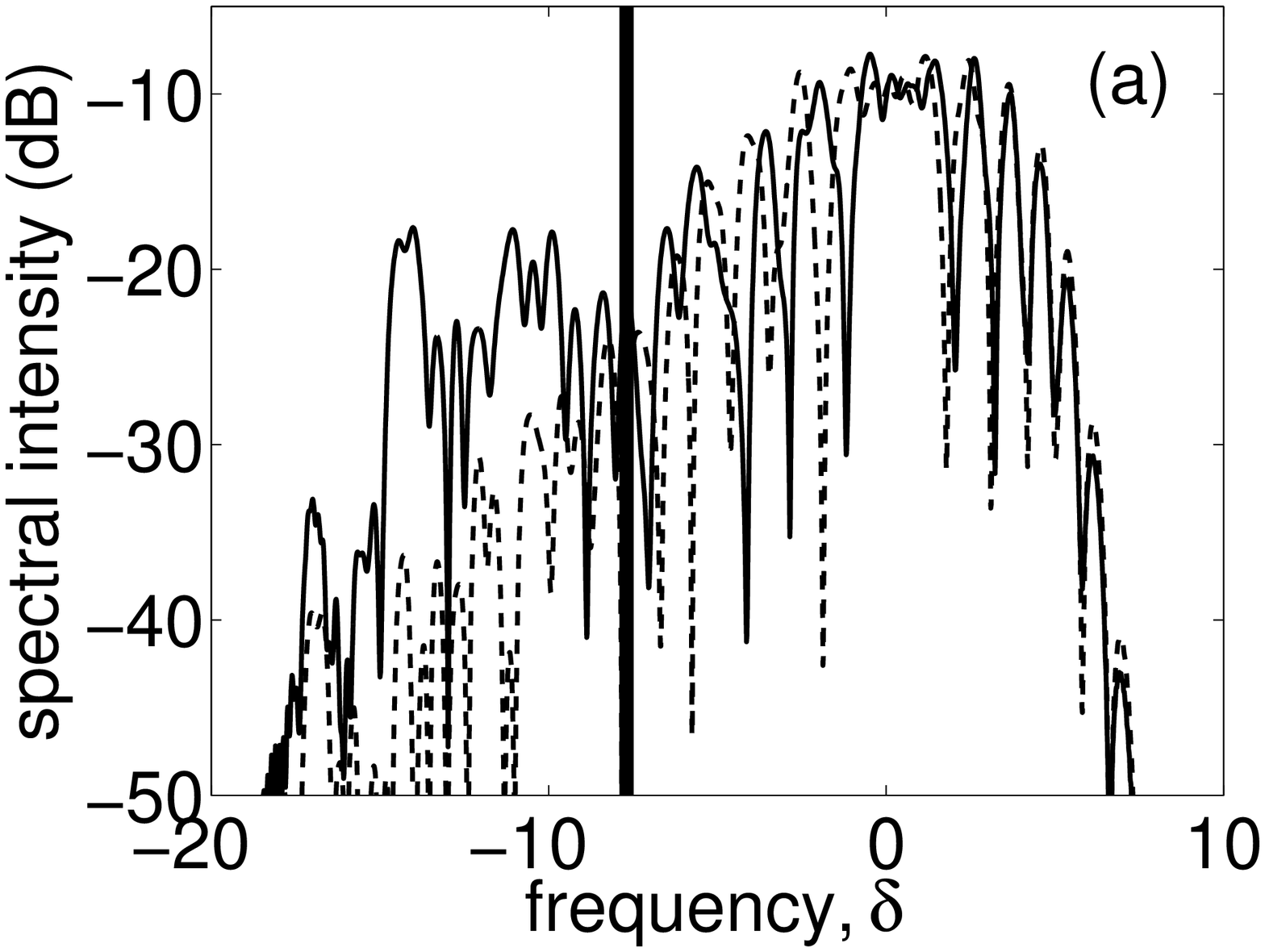}
\includegraphics[width=3.5cm]{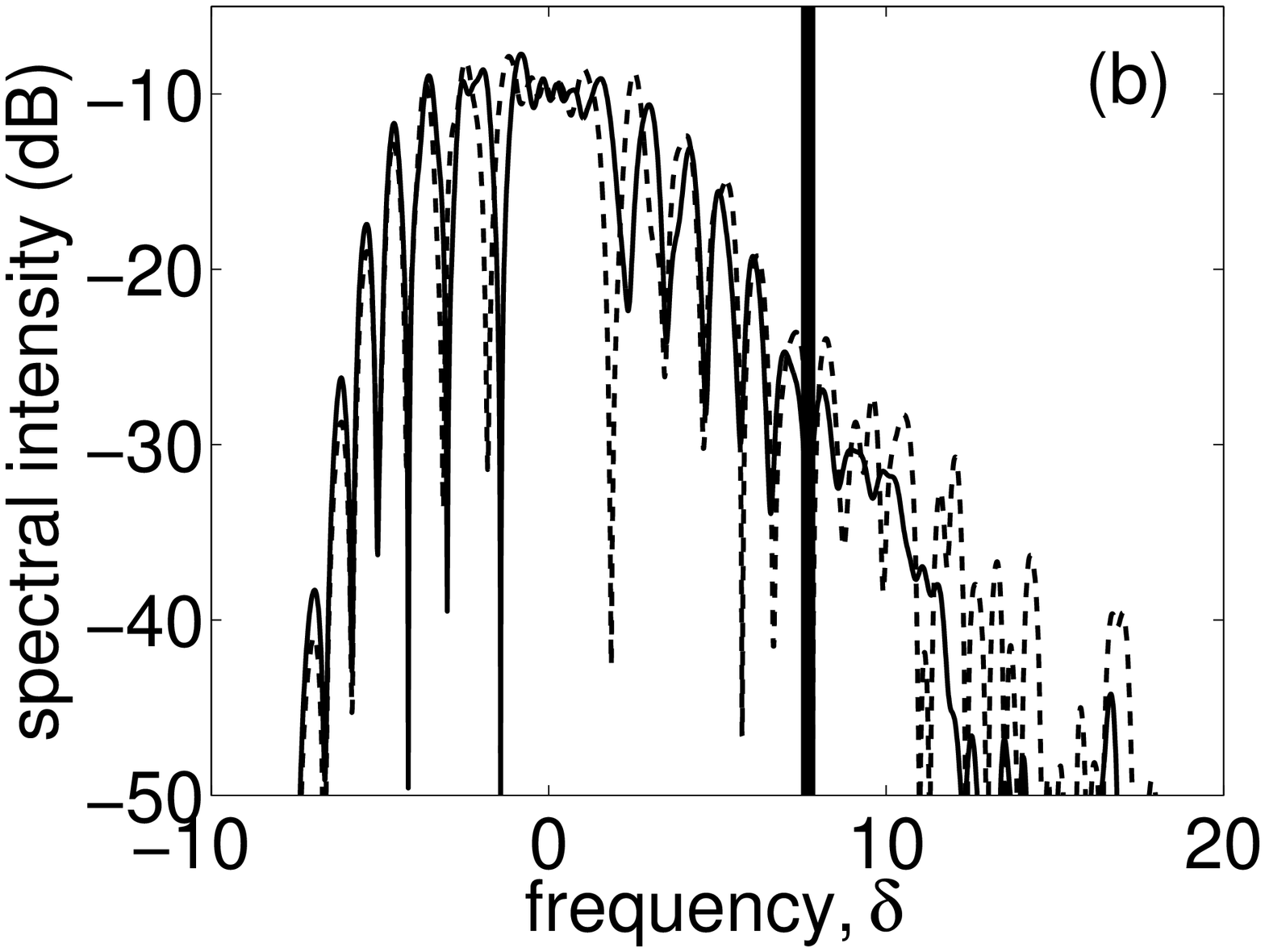}
}
\caption{ }
\label{fig5}\end{figure}

\begin{figure}
\centerline{
\includegraphics[width=3.5cm]{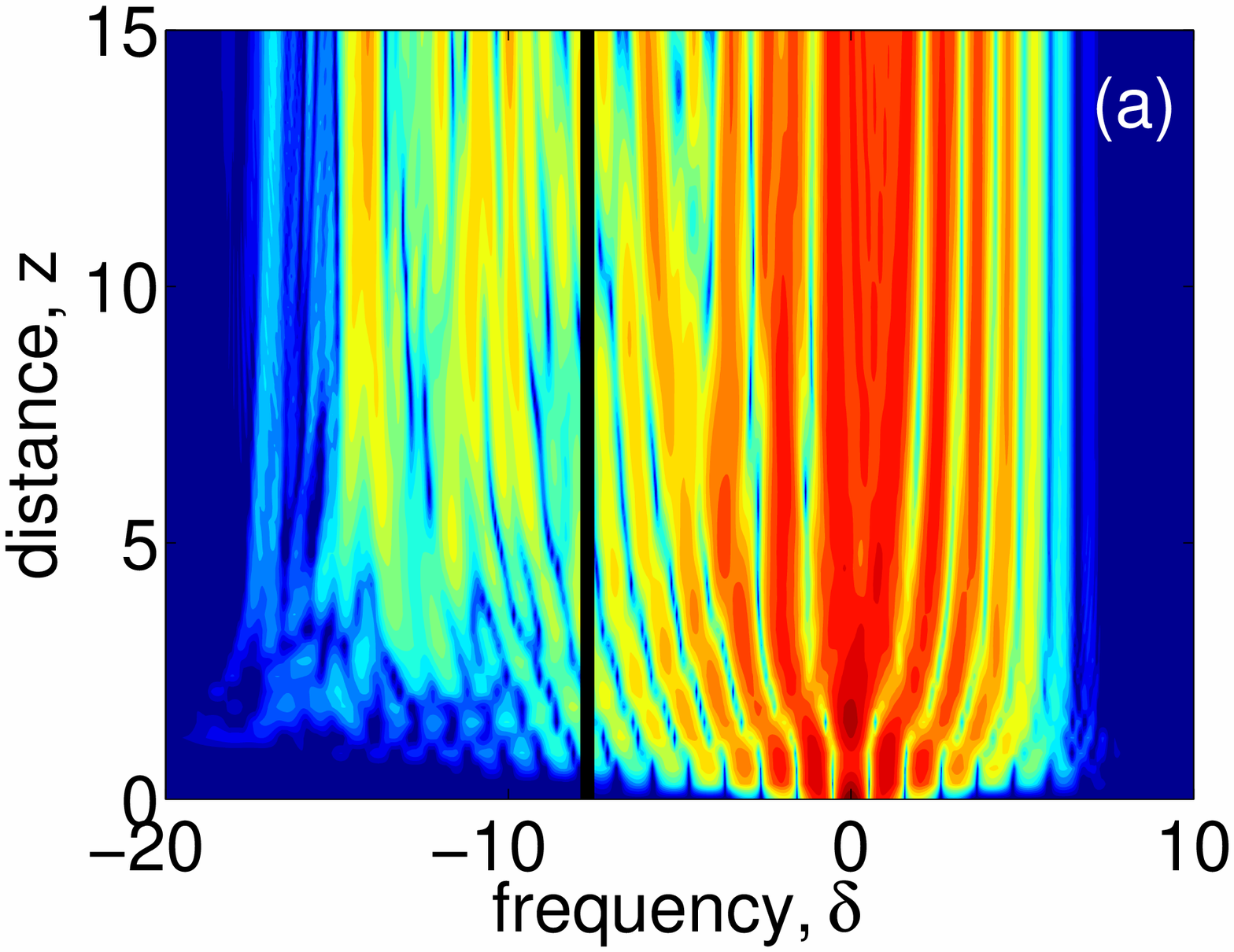}
\includegraphics[width=3.5cm]{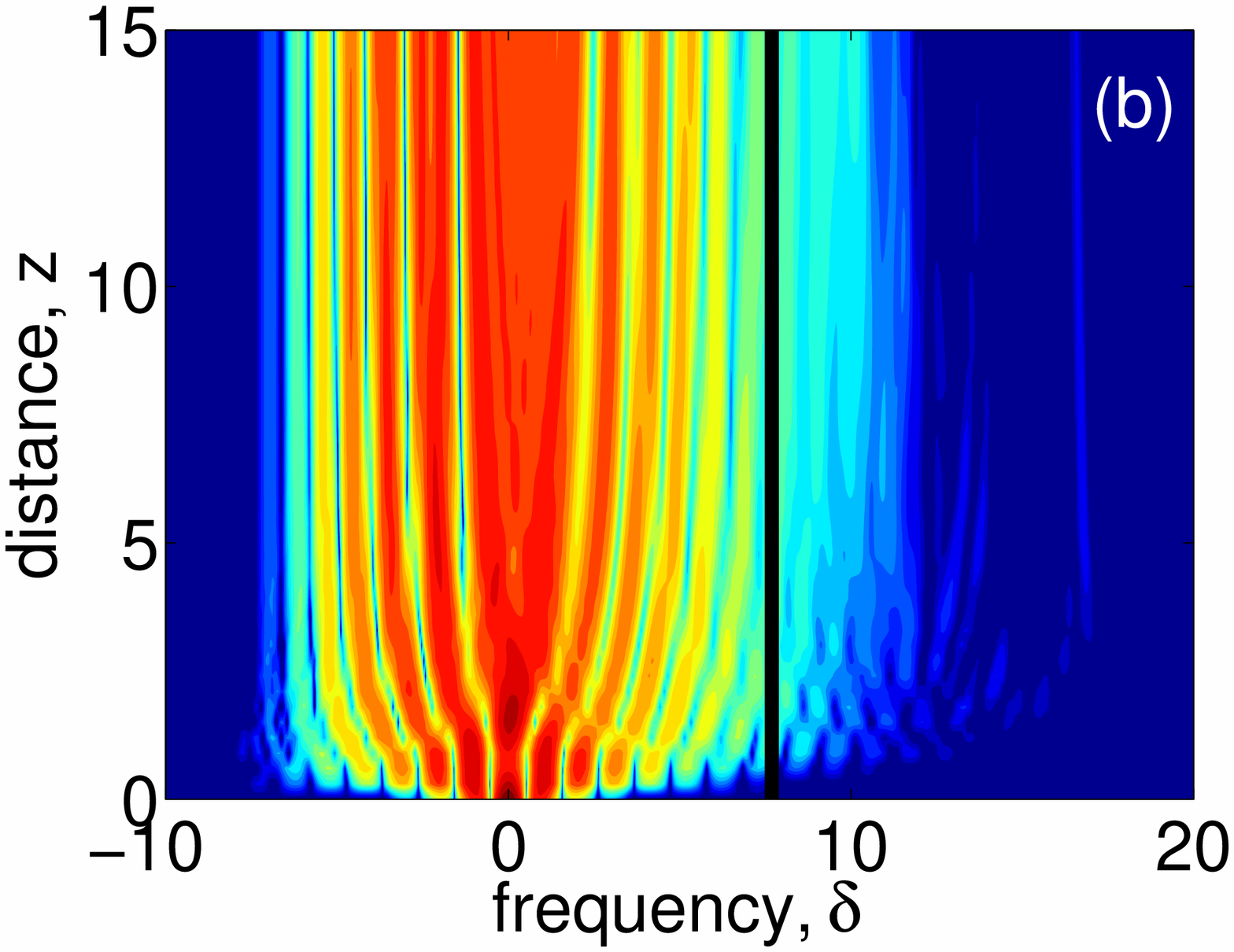}
}
\caption{ }
\label{fig4}\end{figure}

\end{document}